%% file: main.tex
\documentclass[conference]{IEEEtran}
\IEEEoverridecommandlockouts
\usepackage{cite}
\usepackage{amsmath,amssymb,amsfonts}
\usepackage{graphicx}
\usepackage{textcomp}
\usepackage{xcolor}
\usepackage{booktabs}
\usepackage{multirow}
\usepackage{tikz}
\usetikzlibrary{positioning,arrows.meta,fit,backgrounds,calc}
\usepackage[hidelinks]{hyperref}

\newcommand{\code}[1]{\texttt{\small #1}}
\newcommand{\cls}[1]{\texttt{\scriptsize #1}}
\newcommand{\cmark}{\checkmark}
\newcommand{\xmark}{--}

\renewcommand{\baselinestretch}{0.9793}

\begin{document}
\bstctlcite{IEEEexample:BSTcontrol}

\title{The Grid's Own Language: A Common Information Model Layer for LLM Grid Assistants}


\author{\IEEEauthorblockN{Antonio Alc\'antara\textsuperscript{1}, Ioannis Karampinis\textsuperscript{2}, and Spyros Chatzivasileiadis\textsuperscript{2}}
\IEEEauthorblockA{\textsuperscript{1}\textit{CUNEF Universidad}, Madrid, Spain \quad \textsuperscript{2}\textit{Technical University of Denmark (DTU)}, Kgs. Lyngby, Denmark\\
antonio.alcantara@cunef.edu, \{iokar, spchatz\}@dtu.dk}
\thanks{This work was supported by the ERC Starting Grant VeriPhIED, Grant Agreement 949899, funded by the European Research Council.}}

\maketitle

\begin{abstract}
Assistants built on large language models (LLMs) are entering power system operation. Current designs delegate computations to validated solvers, but the agent still infers the type, role and state of each element from its name and tool-specific formats, and errors propagate to every study. The Common Information Model (CIM), the IEC standard in which operators already exchange network models, defines unambiguously what each element is and how it connects. We propose a CIM layer for LLM grid assistants and implement it in CONDUCTOR, an open-source assistant. Both the agent and its power-system studies read the layer. It labels the provenance of each fact (stated in the data, suggested only by a name, or inferred from topology), loads operators' CGMES models directly, and builds in $\leq$1\,s on the 9,241-bus PEGASE network. Deterministic checks compare every answer with solver results and the layer. On 37 tasks about the CIGRE medium-voltage and IEEE 30-bus networks, the agent gets right 97\,\% of the graded facts with the layer, such as a unit's energy source or a switch's normal state, against 55\,\% without it. We argue that a CIM layer should be a default component of grid assistants.
\end{abstract}


\section{Introduction}

AI assistants based on large language models (LLMs) let an operator ask for a power-system study in plain language, such as a security assessment or a corrective dispatch. The prevailing design keeps the LLM away from the numerical computations: it chooses and sets up validated power-system solvers and explains their results~\cite{jin2025gridmind,she2026pfagent,liu2026gridorch}; recent work also checks that its numbers come from those solvers~\cite{mylonas2026governance,shamseldein2026gridmind}. A recent review concludes that trust in such assistants rests as much on correct computations as on valid input data and models~\cite{trustllm2026review}.

Correct numerical results are, therefore, only half of an answer. The other half is if the input data and models are interpreted correctly: which units are solar, which switch ties two feeders, whether a value was measured or assumed. Today's assistants learn this from element names and raw tables. Names, however, are weak evidence, even within a utility~\cite{choi2024egridgpt}. An assistant that counts a diesel unit with a misleading name among the solar units, or takes a normally open tie to be in service, fails at the first step of situational awareness, i.e., perceiving the elements of the system~\cite{endsley1995sa}, and every study built on that perception inherits the error.

The power industry already has a standard for this. The Common Information Model (CIM)~\cite{iec61970_301} is the standard language in which utilities describe their networks, and European system operators exchange their grid models in it, following the Common Grid Model Exchange Standard (CGMES)~\cite{entsoe_cgmes}. The idea of this paper is simple: add a CIM layer, a description of the network in CIM built from its data, and let both the agent and the studies it runs read it, so that, for example, a study examining lower solar output changes exactly the units the layer says are solar. We implement it in CONDUCTOR\footnote{Code for CIM layer, experiments, and run records available at: \url{https://github.com/antonioalcantaramata/CONDUCTOR}}, an open-source assistant, but any LLM assistant that runs power-system studies can be given a CIM layer. Our contributions are:
\begin{enumerate}
\item \textbf{A CIM layer for LLM grid assistants.} Built by deterministic code, not by the LLM, from a solver's data or a CGMES model, it gives each element its CIM class, infers from the topology information not stored in tables (e.g. which parts of the grid an outage cuts off, which tie restores them), and records how each fact is known: stated in the data, suggested only by a name, inferred, or declared by the operator. Both the agent and the studies it runs read it.
\item \textbf{Automated answer checks.} Fixed rules, not a second AI model, check every number against the solver results and every statement about an element against the CIM layer. When a question leaves a study parameter open (e.g. the load power factor), the checks re-run the study under other values and report whether the conclusion changes. In our experiments, the conclusion changed for 1 in 5 re-checked answers.
\item \textbf{An evaluation of what the layer adds.} On 37 questions about the network (e.g., what is a switch's normal state), the agent with no CIM layer gets 55\,\% correct answers against 97\,\% when the agent reads the CIM layer too.
\end{enumerate}

Table~\ref{tab:related} compares this work with prior LLM grid assistants. Prior assistants compute their numbers with solvers, and some check them, but none is told what each element is, nor checks that its answer gets it right. CIM has reached LLMs only as a model to query: in CIMplify Chat, the LLM writes code that queries or edits a CIM model, without seeing the model itself~\cite{anderson2026cimplify}; another method turns parts of a CGMES model into text from which an LLM can answer~\cite{cimgraphrag2026}; and Norway's transmission system operator (TSO), Statnett, is building an LLM access to its CIM models.\footnote{\url{https://github.com/statnett/Talk2PowerSystem}} In these works, the model is only queried, never used in a power-system study. To our knowledge, this is the first CIM layer for LLM grid assistants that records how each fact is known and is read both by the agent and by the studies it runs; CONDUCTOR is also the first assistant to re-run its studies to test whether the conclusion survives other readings.

\begin{table}[t]
\caption{What prior LLM grid assistants and this work provide}
\label{tab:related}
\centering
\footnotesize
\setlength{\tabcolsep}{1.6pt}
\begin{tabular}{@{}lccccc@{}}
\toprule
& Solver & Number & Gover- & CIM & \textbf{This}\\
& agents & check & nance & Q\&A & \textbf{work}\\
& \cite{jin2025gridmind,she2026pfagent} & \cite{shamseldein2026gridmind} & \cite{mylonas2026governance}
& \cite{anderson2026cimplify,cimgraphrag2026} & \\
\midrule
Numbers in the answer checked & (\cmark)$^a$ & (\cmark)$^b$ & (\cmark)$^c$ & \xmark & \cmark\\
Knows what elements are (CIM) & \xmark & \xmark & \xmark & \cmark & \cmark\\
Says how each fact is known & \xmark & \xmark & \xmark & \xmark & \cmark\\
Studies read what elements are & \xmark & \xmark & \xmark & \xmark & \cmark\\
Conclusion re-checked & \xmark & \xmark & \xmark & \xmark & \cmark\\
Operator can add facts & \xmark & \xmark & \xmark & (\cmark)$^d$ & \cmark\\
\bottomrule
\multicolumn{6}{@{}p{\columnwidth}@{}}{\scriptsize \cmark{} yes; (\cmark) in part; \xmark{} no. $^a$ Offline, against reference values. $^b$ Whether any tool ran. $^c$ Its governance layer writes them. $^d$ Via generated code~\cite{anderson2026cimplify}.}\\
\end{tabular}
\end{table}

\section{The Common Information Model (CIM)}

CIM is a shared vocabulary for power systems, maintained by the International Electrotechnical Commission (IEC): it defines what each element of a network is, what properties it has and how it relates to the others, so that tools reading CIM understand a network the same way. A photovoltaic (PV) plant is a \code{PowerElectronicsConnection} with a photovoltaic unit, regardless of its name. A switch carries its normal state (\code{normalOpen}). A bus belongs to a \code{VoltageLevel} in a \code{Substation}, or to a \code{Feeder}. A complete CIM model needs no provenance, i.e., no record of how each fact is known: its classes state each fact. Assistants, however, usually receive a network in a solver's own format, such as pandapower's tables~\cite{thurner2018pandapower} or a MATPOWER case, or as a CIM model with generic classes; there a unit's type may appear only in its name, or nowhere. Converted to CIM, a class guessed from a name would look like one stated in the data; our CIM layer therefore records the provenance of each fact. And even a complete model does not store what an outage cuts off or which tie restores it: the layer infers these from the topology.


\section{Integrating the CIM Layer into CONDUCTOR}

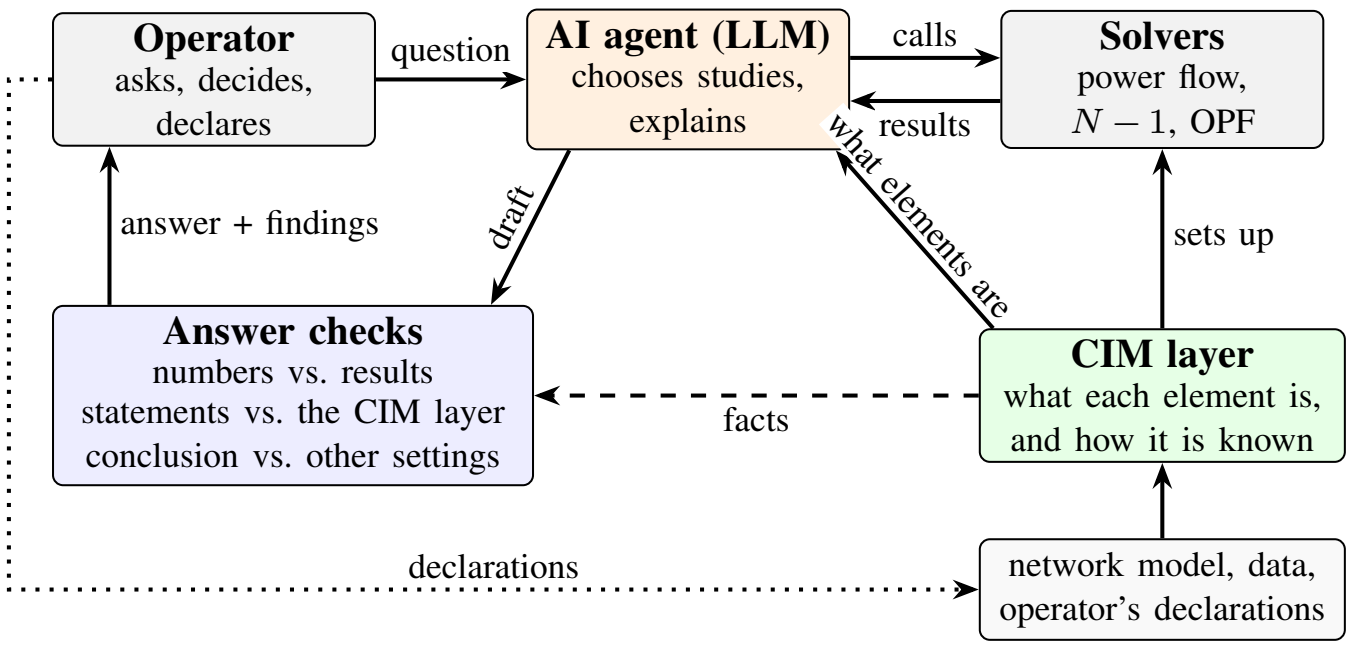
\begin{figure}[t]
\centering
\resizebox{\columnwidth}{!}{%
\begin{tikzpicture}[
  box/.style={draw, rounded corners=2pt, align=center, font=\footnotesize, inner sep=2pt, minimum height=9mm,
              text width=21mm},
  arr/.style={-{Stealth[length=1.8mm]}, thick},
  lab/.style={font=\scriptsize, fill=white, inner sep=0.8pt}]
\node[box, fill=gray!10]   (op)   at (0, 0)    {\textbf{Operator}\\\scriptsize asks, decides,\\declares};
\node[box, fill=orange!12] (llm)  at (3.3, 0)  {\textbf{AI agent (LLM)}\\\scriptsize chooses studies,\\explains};
\node[box, fill=gray!10]   (solv) at (6.6, 0)  {\textbf{Solvers}\\\scriptsize power flow,\\$N\!-\!1$, OPF};
\node[box, fill=blue!7, text width=32mm] (chk) at (0.55, -2.2) {\textbf{Answer checks}\\
  \scriptsize numbers vs.\ results\\ statements vs.\ the CIM layer\\ conclusion vs.\ other settings};
\node[box, fill=green!10, text width=24mm] (sem) at (6.6, -2.2) {\textbf{CIM layer}\\\scriptsize what each element is,\\and how it is known};
\node[box, fill=gray!5, text width=24mm, minimum height=7mm] (net) at (6.6, -3.55)
  {\scriptsize network model, data,\\operator's declarations};
\draw[arr] (op) -- node[lab, above=1pt]{question} (llm);
\draw[arr] ([yshift=1.5mm]llm.east) -- node[lab, above=1pt]{calls} ([yshift=1.5mm]solv.west);
\draw[arr] ([yshift=-1.5mm]solv.west) -- node[lab, below=1pt]{results} ([yshift=-1.5mm]llm.east);
\draw[arr] (sem) -- node[lab, right=1pt]{sets up} (solv);
\draw[arr] ([xshift=1mm]sem.north west) -- node[lab, sloped, above=0.5pt, pos=0.56]{what elements are} ([xshift=-1mm]llm.south east);
\draw[arr] ([xshift=3mm]llm.south west) -- node[lab, sloped, above=1pt]{draft} ([xshift=-3mm]chk.north east);
\draw[arr, dashed] (sem.west) -- node[lab, below=1pt]{facts} (chk.east);
\draw[arr] ([xshift=4mm]chk.north west) -- node[lab, right=1pt]{answer + findings} ([xshift=4mm]op.south west);
\draw[arr] (net) -- (sem);
\draw[arr, dotted] (op.west) -- ++(-0.3, 0) |- node[lab, pos=0.75, above=1pt]{declarations} (net.west);
\end{tikzpicture}}
\caption{CONDUCTOR. The CIM layer tells the agent and the solvers what each element is; every answer reaches the operator with the checks' findings.}
\label{fig:arch}
\end{figure}

\subsection{The CIM layer}\label{sec:layer}

CONDUCTOR builds the CIM layer from whatever network is loaded: from a CGMES model, it reads each element's CIM class; from a solver's own data, such as pandapower's tables or a MATPOWER case, it assigns one (Table~\ref{tab:example}). The layer describes each element through facts, statements such as ``PV~9 is solar'', of the following three \emph{fact types}: (i)~\emph{Attributes} that the data hold but the agent never sees, since it receives only names, connections and study results: what each bus, branch, unit, load and switch is, such as a unit's energy source or a switch's normal state, each with a unique, persistent identifier. (ii)~\emph{Relations} that no data table stores, which the layer derives from the topology: what an outage disconnects, which feeder supplies a load. (iii)~\emph{Provenance}, i.e., how each fact is known: \emph{stated} in the data, \emph{named} (only the name suggests it), \emph{inferred} from the topology, \emph{declared} by the operator, or \emph{unknown}. Each numerical value, such as a load's reactive power, also records its origin: measured, or derived from an assumed power factor.

The operator can also add what the data lack (``unit 12 is a wind farm''): the declaration becomes a fact marked \emph{declared}, and every later study uses it. The layer can be written out as a CIM model, with the provenance of each fact stored in an extension of the standard, so that other CIM tools can read it.

\begin{table}[t]
\caption{What the CIM layer says about 3 elements of CIGRE MV}
\label{tab:example}
\centering
\footnotesize
\setlength{\tabcolsep}{3pt}
\begin{tabular}{@{}p{0.45\columnwidth}p{0.52\columnwidth}@{}}
\toprule
Element and CIM class & What the layer says, and how it is known\\
\midrule
Solar Farm North\newline\cls{SynchronousMachine,}\newline\cls{ThermalGeneratingUnit} & Thermal, \emph{stated}: its type
in the data is ``CHP diesel''. The name is not evidence.\\
Bus 9\newline\cls{TopologicalNode} & 20\,kV, \emph{stated}; supplied by the feeder of Line 1-2, \emph{inferred} from
the topology.\\
S1\newline\cls{LoadBreakSwitch} & Normally open, \emph{stated}; the tie between the two feeders, \emph{inferred} from
what it switches.\\
\bottomrule
\end{tabular}
\end{table}

\subsection{Architecture}

Fig.~\ref{fig:arch} shows the architecture of CONDUCTOR. The agent is an LLM with about 30 tools, functions it can call to run a study or look up an element; it chooses the studies and explains their results, but is instructed never to compute a number it reports. Instead, every number comes from a \emph{study}, a computation the agent calls and a solver runs: pandapower runs the power flows and contingency ($N\!-\!1$) screens, and IPOPT the optimal power flow (OPF) and corrective dispatch.

The CIM layer is read in three ways. First, the agent queries it, for the elements a user question concerns: each query returns its facts and provenance, which enter the agent's context. Second, the studies query it: asked to halve wind and solar output, a study changes exactly the units the CIM layer says are wind or solar. Third, the answer checks (Section~\ref{sec:answerchecks}) compare the agent's statements with it.

\subsection{Checking the agent's answers}\label{sec:answerchecks}

Before an answer reaches the user, three checks, fixed rules rather than a second AI model, compare the answer with what is known; what they flag is shown to the user below the answer.

\emph{Numbers.} Every number in the answer is compared with the values it came from (e.g. the solver results, the study parameters or the question), at the precision it is written (e.g., 0.94 matches 0.942). A number is flagged if (a) it matches no such value, (b) it matches a value of another element than the one its sentence is about, or (c) the study it comes from contradicts it, e.g. an insecure result called secure. A claim worded more strongly than its evidence supports is also flagged; for instance, stating that no feasible dispatch exists when the optimiser only failed to find one.

\emph{Statements about elements.} What the answer says an element is (source, voltage, fuel type, switch state, measured or not) is compared with the CIM layer. A statement the layer contradicts is a failure; one it does not support is a warning.

\emph{Conclusions.} A question often leaves some study parameters open: the load power factor, the loading limit, the voltage band, etc. After the answer is written, the studies on which the conclusion rests are re-run under the alternatives the question left open, and their verdicts (e.g., secure or insecure) are compared. The conclusion can \emph{hold}, \emph{shift} (same verdict, but a numerical value moves by over 10\,\%) or \emph{change}; a change is reported with the answer.


\section{Case Study}\label{sec:case}

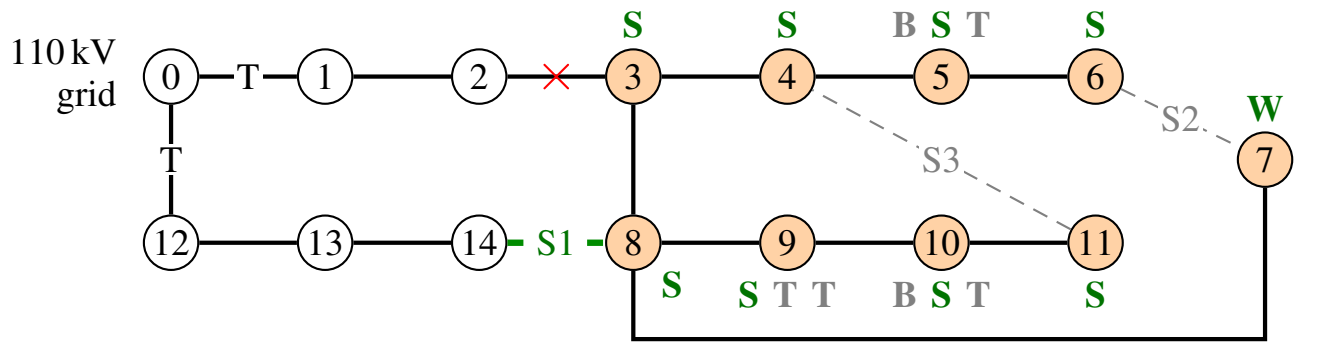
\begin{figure}[t]
\centering
\resizebox{\columnwidth}{!}{\input{fig_case.tex}}
\caption{The CIGRE MV network, drawn from the CIM layer. If Line~2-3 trips (\textcolor{red}{$\times$}), the layer infers from the topology that \figcasefacts; dashed: normally open. Unit sources as stated: \textbf{S} solar, \textbf{W} wind, \textbf{T} thermal, \textbf{B} battery (green: changed by a wind and solar scenario).}

\label{fig:case}
\end{figure}

We use two systems. First, the CIGRE medium-voltage (MV) benchmark~\cite{strunz2014cigre} as implemented in pandapower (Fig.~\ref{fig:case}): a 110/20\,kV substation feeding two radial feeders, with 15 buses, 18 loads, eight PV units, a wind turbine, a diesel combined heat and power (CHP) unit, fuel cells and two batteries, each with its type stated in the data, and eight switches, three of them normally open. A synthetic week (1--7 January) at 15-minute resolution gives each load's active power, from a residential daily profile, and reactive power at a 0.97 power factor; it also gives each unit's output from its stated type: the sun for PV, the wind for the turbine, flat for the others. At the evening peaks, voltages along the longer feeder fall below the 0.95\,p.u.\ limit. Second, we use as a contrast the IEEE 30-bus case from the Power Grid Library (PGLib)~\cite{babaeinejadsarookolaee2019pglib}, whose data give each generating unit only the bus it is connected to, its power limits and a cost, with no name or type, so ``unknown'' is often the right answer.

One logged session shows how the CIM layer and the checks work. Asked \emph{``If Line 2-3 trips, which loads lose supply, and which switch could restore them?''}, the agent answers with the facts of Fig.~\ref{fig:case}, read from the CIM layer rather than guessed. Because the layer infers S1's role from the topology alone, without a power flow, the answer warns that closing S1 is not a verified switching plan. 
Asked \emph{``Is the grid secure at the evening peak on 3 January?''}, the agent answers no: 9 buses are below 0.95\,p.u., the lowest is Bus~11 at 0.942\,p.u.; the checks trace every number in the answer to the study or its inputs. Since the question does not state a load power factor, the checks re-run the study under other readings: at 1.00 instead of the default 0.97 it finds no violation, and the answer gives both results and asks which was meant.


\section{Experiments}\label{sec:exp}
%
We wrote 88 tasks. Each is one question, or a question with a follow-up (e.g., the operator first declares a fact, then asks). Each task has criteria, points that its answer, or the study behind it, must state or avoid; the agent is never shown them, and each is checked automatically against the network or the study's own record, never by a model. Of these, 37 tasks ask about the network, broken down in Table~\ref{tab:ontology}. Most are plain look-ups (what an element is, which switches are normally open, whether a value is measured); the rest test harder cases: \emph{misleading names}, on copies of the networks in which we renamed units (e.g., a CHP unit called ``Solar Farm North'' must not count as solar); \emph{changes to wind and solar}, whether a study that halves only their output changes the right units; \emph{combining several facts}, such as what an outage cuts off and which tie restores it; and \emph{consistency}, whether the answer stays right when the same question is reworded or asked after a declared fact. The other 51 tasks ask for studies: what-ifs, investigations, corrective options, clarifications and reports. Every run uses the same LLM (OpenAI's gpt-6-luna, medium reasoning effort), with the criteria fixed beforehand. We report the shares of criteria met and of runs meeting all of them. To compare a variant with the full system, we take, for each task, the difference in the share of its criteria met, and average it over tasks; its 95\,\% interval comes from resampling the tasks (bootstrap), and we call a difference significant when the interval excludes zero.

\subsection{What does the CIM layer give the agent and the studies?}\label{sec:grounding}

Experiment~1 runs the 37 network tasks 5 times in 3 variants (Table~\ref{tab:ontology}): the full system (CIM layer used by both agent and studies); the CIM layer hidden from the agent but used by the studies; and no layer at all. Hiding the layer from the agent removes the layer's two tools (one describes the network, one records declarations) and the facts it adds to look-ups: the agent still sees every element's name, connections and study results, but not types, switches or where values come from. The LLM and its instructions are the same in every variant, except one rule, removed in the two variants where the agent does not use the layer: take what an element is from the layer's tools, say how each fact is known, and when the network does not state a fact, say so rather than guess it from the name. Without the CIM layer, the share of criteria met drops from 97\,\% to 65\,\% when only the studies use it, and to 55\,\% with no layer; both drops are significant.

\begin{table}[t]
\caption{Experiment 1: criteria met (\%) on questions about the network}
\label{tab:ontology}
\centering
\footnotesize
\setlength{\tabcolsep}{4pt}
\begin{tabular}{@{}lccc@{}}
\toprule
& \multicolumn{3}{c}{CIM layer used by:}\\
& agent and & studies & \\
Task family (tasks) & studies & only & neither\\
\midrule
What elements are, CIGRE MV (5) & 90 & 77 & 73\\
What elements are, IEEE 30 (5) & 100 & 77 & 73\\
Misleading names (5) & 100 & 24 & 20\\
Changes to wind and solar (4) & 100 & 100 & 42\\
Combining several facts (4) & 96 & 74 & 68\\
Switches (4) & 100 & 22 & 22\\
Where values come from (6) & 92 & 55 & 55\\
Consistency (4) & 98 & 100 & 100\\
\midrule
All criteria (37 tasks) & \textbf{97} & 65 & 55\\
Runs meeting every criterion & \textbf{95} & 54 & 42\\ \bottomrule 
\end{tabular}
\end{table}

\emph{Without the CIM layer, the agent misses the facts it cannot see.} It trusts names: asked which units are solar, its first run answered ``There are 10 solar units: PV 3, \ldots, PV 11, Solar Farm North, and PV Park East'', counting the diesel CHP unit and a fuel cell as PV, misled by their names; with the layer, every run on misleading names was correct. Without the CIM layer, the agent cannot name the normally open switches or the tie between the feeders, nor tell a measured value from a value derived from an assumed power factor, so it either says it cannot tell or guesses. Each miss belongs to one of the three fact types of Section~\ref{sec:layer}: (i)~attributes, (ii)~relations or (iii)~provenance. The gain comes not from a better model but from better facts, which the CIM layer supplies.

\emph{Without any CIM layer, the studies change the wrong units.} While the studies read the layer, a 50\,\% cut in wind and solar changes exactly the wind and PV units, even if the agent cannot see the layer (100 in both columns of Table~\ref{tab:ontology}); with no layer, it also cuts the CHP unit and fuel cells, and declarations change nothing.

\emph{Raw data tables instead of the CIM layer.} In a fourth variant, the agent reads the network's raw data tables (pandapower's own, one row per bus, line, unit or switch) instead of the CIM layer, while the studies still use it. The tables hold fact type~(i), i.e., attributes such as unit types and switch states, so names rarely mislead the agent; but they hold neither fact type~(ii), relations, nor (iii), provenance. In no run did the agent name the feeder that supplies a load or the tie that restores supply after an outage, and in most it missed that a PV unit's output is not measured on its own: the data give each bus's total production, which is split among the units there by capacity. The agent met 84\,\% of the criteria, above the 65\,\% with the layer hidden but below the 97\,\% with it, and was slower (23\,s against 9\,s). Raw tables are thus no substitute for the CIM layer, nor do they scale: written out as text, they take about 200 tokens (units an LLM reads, $\approx4$ characters each) per bus, so the 9,241-bus PEGASE network would need 1.9 million, far beyond what an LLM can read.

\subsection{What does each part of CONDUCTOR add?}\label{sec:components}

Experiment~2 measures what each part of CONDUCTOR adds. It runs all 88 tasks (37 about the network, 51 asking for studies) three times in each of eight variants (2,112 runs, all completed; Table~\ref{tab:components}): the full system; six variants that each remove one part and keep the rest; and \emph{solvers only}, an agent that keeps every solver tool but has no CIM layer and no re-run, as prior assistants are built~\cite{jin2025gridmind,she2026pfagent}. The six parts are 1)~the CIM layer for the agent; 2)~the re-run of the parameters a question left open (Section~\ref{sec:answerchecks}); two instructions to the agent: 3)~the \emph{routing rules}, which send each type of question to its tool, and 4)~the \emph{evidence wording}, which asks the agent to word each claim by what backs it; 5)~asking first, when the operator has left something open, such as which line; and 6)~the investigation tools, which trace why a violation occurs, which units and loads drive it, and what would prevent it.

\begin{table}[t]
\caption{Experiment 2: criteria met (\%) with one part removed}
\label{tab:components}
\centering
\footnotesize
\setlength{\tabcolsep}{2.1pt}
\begin{tabular}{@{}lrcrr@{}}
\toprule
& \multicolumn{2}{c}{All 88 tasks} & & \\
\cmidrule(lr){2-3}
System & \% & $\Delta^a$ [95\,\% CI] & Network & Studies\\
\midrule
Full system & 97.1 & -- & 99.0 & 96.3\\
1)~w/o CIM layer for agent & 87.8 & $-13.2$ [$-20.3$, $-6.5$] & 64.1 & 98.6\\
2)~w/o re-run of parameters & 95.7 & $-1.1$ [$-3.1$, $+0.8$] & 95.9 & 95.6\\
3)~w/o routing rules & 95.0 & $-2.3$ [$-4.6$, $-0.2$] & 97.4 & 93.9\\
4)~w/o evidence wording & 97.6 & $+0.1$ [$-2.1$, $+2.1$] & 96.9 & 97.9\\
5)~w/o asking first & 97.3 & $+0.7$ [$-1.7$, $+3.3$] & 97.9 & 97.0\\
6)~w/o investigation tools & 84.8 & $-12.3$ [$-18.3$, $-6.8$] & 95.9 & 79.7\\
Solvers only & 84.6 & $-17.4$ [$-25.2$, $-10.0$] & 53.8 & 98.6\\
\bottomrule
\multicolumn{5}{@{}p{0.98\columnwidth}@{}}{\scriptsize \%: share of all criteria met. $^a$ Difference from the full system in points, per task, averaged over tasks (not the difference of the \% column), with a bootstrap interval. Network: the 37 tasks about the network; Studies: the 51 others.}\\
\end{tabular}
\end{table}

The solvers-only agent meets 85\,\% of the criteria against 97\,\%, all lost on questions about the network (54\,\% against 99\,\%). The other parts matter less. The re-run (2) barely changes the score, because it only adds a note to the answer, yet it matters: in 21 of the 103 answers it checked (one in five), a different value of a parameter the question left open, such as the load power factor, changed the conclusion. The routing rules (3) give a small gain, the evidence wording (4) none, and asking first (5) matters only on clarification tasks. Removing the investigation tools (6) lowers the score by design, as most criteria of the investigation tasks ask whether the agent investigated.

\subsection{Do the answer checks catch errors?}\label{sec:checks}

\begin{table}[t]
\caption{Errors put into recorded answers, and the share flagged}
\label{tab:injection}
\centering
\footnotesize
\setlength{\tabcolsep}{2pt}
\begin{tabular}{@{}lr@{\hspace{8pt}}rr@{}}
\toprule
Error put into the answer & Injected & Flagged (\%) & 95\,\% CI\\
\midrule
\multicolumn{4}{@{}l}{\emph{Numbers}}\\
\quad A number changed by a few \% & 1,424$^a$ & 93.5 & 92.1--94.6\\
\quad Another element's number & 59 & 100 & 93.9--100\\
\quad An insecure result called secure & 198 & 100 & 98.1--100\\
\quad A claim beyond its evidence & 262 & 98.1$^b$ & 95.6--99.2\\
\multicolumn{4}{@{}l}{\emph{Statements about elements, against the CIM layer}}\\
\quad A unit called the wrong source & 427 & 100 & 99.1--100\\
\quad A source the network does not state & 642 & 100 & 99.4--100\\
\quad A bus at a voltage it does not have & 1,748 & 100 & 99.8--100\\
\quad A line called a transformer & 514 & 99.6 & 98.6--99.9\\
\quad An assumed value called measured & 1,507 & 100 & 99.7--100\\
\quad An element the network does not have & 2,457 & 99.5 & 99.1--99.7\\
\midrule
True statements flagged (controls) & 3,534 & 0 & 0--0.1\\
\bottomrule
\multicolumn{4}{@{}p{\columnwidth}@{}}{\scriptsize $^a$ A further 716 changed numbers equal another result of the same answer and are not scored. $^b$ The 5 missed were in follow-ups relying on the previous turn's study.}\\
\end{tabular}
\end{table}

Prior checks in LLM grid assistants verify only the numbers in an answer~\cite{mylonas2026governance,shamseldein2026gridmind}. With the CIM layer, CONDUCTOR's checks also verify what an answer says about the network, such as a unit's source or whether a value was measured (lower half of Table~\ref{tab:injection}). To test the checks, we put one known error at a time, such as a voltage changed by a few percent or a PV unit called wind, into copies of the 2,717 answers recorded in Experiments~1 and~2. An error counts as caught only if the check flags the number or sentence that contains it. To test for false alarms, we also checked 3,534 true statements.

\emph{Injected errors.} Every kind of error is caught in 93--100\,\% of cases, and those in statements about elements in at least 99.5\,\%; no true statement is flagged. This is a best case: each injected error was written in a form the checks can read, which the agent's own errors need not be, as we show next.

\emph{The agent's own errors} are a harder test. Misleading names fooled only \emph{the agent without the CIM layer}: in 30 answers in Experiment~1, it trusted a name that the data contradict, such as counting ``Solar Farm North'', a diesel CHP unit, as solar; with the layer, every answer on misleading names was correct. The checks run after the answer is written and always read the CIM layer, even in the variants where the agent could not: they flagged 12 of these 30 answers and missed the other 18, worded in ways the checks do not recognise. Across all recorded answers, most warnings flagged no real error but asked for the calculation behind a correct number. In short, the CIM layer prevents these errors and lets the checks verify what an answer says about the network, not only its numbers.

\subsection{Real CIM models and scalability}\label{sec:real}

So far, our experiments used two small benchmarks, because their facts are known, so every answer can be graded. A system operator's own model is larger, arrives in CIM format, and is often modelled down to each breaker. We therefore tested the layer alone, without the LLM: whether it reads such a model, finds the structure of a realistic distribution grid, and scales to a transmission network. SmallGrid, a CGMES model that ENTSO-E publishes to test whether tools read the format correctly, has 1,225 buses and 1,107 switches; from it, the layer takes every identifier, substation, voltage, branch kind, switch type and unit type as stated, in $<$1\,s. On MV Oberrhein, a 20\,kV grid with 179 buses and 322 switches, it builds in 0.03\,s, finds its four feeders and classifies its six normally open switches: five tie two feeders and one closes a loop within a feeder. When the first line of a feeder trips, it identifies, depending on the feeder, the 32 to 41 buses cut off and the 1 to 4 ties that could restore them. On the 9,241-bus PEGASE case, it builds in $\leq$1\,s, and an answer is checked in $<$10\,ms.

\section{Conclusion}

Despite LLM grid assistants already computing with validated solvers, they still read the network through element names, which leads to wrong facts in their answers. This paper introduces a CIM layer, built automatically from a solver's data or a CGMES model, that gives the agent correct and traceable facts about the network and extends CIM with how each fact is known. 
With the layer read both by the agent and by the studies it runs, the agent interprets correctly 97\,\% of the graded facts about the network, against 84\,\% with the raw data tables and at most 65\,\% without the layer. Agents thus need not only accurate network data but also the relations and provenance drawn from them; the CIM layer offers both.

We also introduce automated answer checks, fixed rules rather than an AI model, that compare every answer with the solver results and the CIM layer, and re-run the studies when a question leaves parameters open. They caught 93--100\,\% of the errors we injected, and, thanks to the layer, at least 99.5\,\% of false statements about network elements; the re-runs changed one re-checked conclusion in five. Because operators already keep their networks in CIM, the layer adds no modelling work: it loads their CGMES models directly and builds in $\leq$1\,s on the 9,241-bus PEGASE network. We argue that all future LLM agents for power grids should include a CIM layer by default.


\section*{Acknowledgment}
The authors thank Iason Gram, Data Scientist at Energinet (Danish TSO), whose suggestion led to the CIM layer. Claude Opus 5.5 (Anthropic) was used to draft/rewrite parts of this text and of the software; the authors reviewed all of it and take responsibility for it.

\bibliographystyle{IEEEtran}
\bibliography{references}

\end{document}

%% file: fig_case.tex
\gdef\figcasefacts{Buses 3--11 (shaded; 11 loads, 4.3\,MW nominal) lose supply, and the normally open tie S1 on Line~14-8 could restore them}
\begin{tikzpicture}[x=10.2mm, y=11mm, font=\scriptsize,
  bus/.style={circle, draw, inner sep=0pt, minimum size=3.6mm, fill=white},
  cut/.style={fill=orange!35},
  br/.style={thick}, nop/.style={densely dashed, gray},
  tag/.style={inner sep=0.6pt, font=\scriptsize\bfseries},
  ren/.style={tag, text=green!45!black}, oth/.style={tag, text=gray}]
\node[bus] (b0) at (-1,2) {0};
\node[bus] (b1) at (0,2) {1};
\node[bus] (b2) at (1,2) {2};
\node[bus, cut] (b3) at (2,2) {3};
\node[bus, cut] (b4) at (3,2) {4};
\node[bus, cut] (b5) at (4,2) {5};
\node[bus, cut] (b6) at (5,2) {6};
\node[bus] (b12) at (-1,1) {12};
\node[bus] (b13) at (0,1) {13};
\node[bus] (b14) at (1,1) {14};
\node[bus, cut] (b8) at (2,1) {8};
\node[bus, cut] (b9) at (3,1) {9};
\node[bus, cut] (b10) at (4,1) {10};
\node[bus, cut] (b11) at (5,1) {11};
\node[bus, cut] (b7) at (6.1,1.5) {7};
\node[left=0.5mm of b0, align=right] {110\,kV\\grid};
\draw[br] (b1) -- (b2);
\draw[br] (b2) -- node[red, font=\normalsize\bfseries, inner sep=0pt] {$\times$} (b3);
\draw[br] (b3) -- (b4);
\draw[br] (b4) -- (b5);
\draw[br] (b5) -- (b6);
\draw[br] (b7) |- (2,0.42) -- (b8);
\draw[br] (b8) -- (b9);
\draw[br] (b9) -- (b10);
\draw[br] (b10) -- (b11);
\draw[br] (b3) -- (b8);
\draw[br] (b12) -- (b13);
\draw[br] (b13) -- (b14);
\draw[nop] (b6) -- (b7);
\path (b6) -- node[gray, fill=white, inner sep=0.5pt, font=\scriptsize] {S2} (b7);
\draw[nop] (b11) -- (b4);
\path (b11) -- node[gray, fill=white, inner sep=0.5pt, font=\scriptsize] {S3} (b4);
\draw[nop, very thick, green!55!black] (b14) -- (b8);
\path (b14) -- node[green!45!black, fill=white, inner sep=0.5pt, font=\scriptsize] {S1} (b8);
\draw[br] (b0) -- (b1);
\path (b0) -- node[fill=white, inner sep=0.5pt, font=\scriptsize] {T} (b1);
\draw[br] (b0) -- (b12);
\path (b0) -- node[fill=white, inner sep=0.5pt, font=\scriptsize] {T} (b12);
\node[tag, above=0.4mm of b3] {\textcolor{green!45!black}{S}};
\node[tag, above=0.4mm of b4] {\textcolor{green!45!black}{S}};
\node[tag, above=0.4mm of b5] {\textcolor{gray}{B} \textcolor{green!45!black}{S} \textcolor{gray}{T}};
\node[tag, above=0.4mm of b6] {\textcolor{green!45!black}{S}};
\node[tag, below right=0.4mm of b8] {\textcolor{green!45!black}{S}};
\node[tag, below=0.4mm of b9] {\textcolor{green!45!black}{S} \textcolor{gray}{T} \textcolor{gray}{T}};
\node[tag, below=0.4mm of b10] {\textcolor{gray}{B} \textcolor{green!45!black}{S} \textcolor{gray}{T}};
\node[tag, below=0.4mm of b11] {\textcolor{green!45!black}{S}};
\node[tag, above=0.4mm of b7] {\textcolor{green!45!black}{W}};
\end{tikzpicture}